\documentclass[sigconf]{acmart}
\acmConference[LLM4EVAL at WSDM '25]{March 2025}{March 2025, Hannover, Germany}

\usepackage{algorithmic}
\usepackage{graphicx}
\usepackage{textcomp}
\usepackage{multirow}
\usepackage{xcolor}
\usepackage{hyperref}
\def\BibTeX{{\rm B\kern-.05em{\sc i\kern-.025em b}\kern-.08em
    T\kern-.1667em\lower.7ex\hbox{E}\kern-.125emX}}

\setcopyright{acmcopyright}
\makeatletter
\renewcommand\@copyrightpermission{%
  \begin{minipage}{\columnwidth}\raggedright\footnotesize
    © 2025 Fitte-Rey et al. This is an open access article distributed under the terms of the  
    Creative Commons Attribution 4.0 International License  
    (http://creativecommons.org/licenses/by/4.0/).
  \end{minipage}%
}

\acmYear{2025}
\copyrightyear{2025}
\acmDOI{}
\acmISBN{}

\begin{document}

\title{Augmented Relevance Datasets with Fine-Tuned Small LLMs}

\author{Quentin Fitte-Rey}
\email{quentinfitterey@hotmail.com}
\affiliation{%
  \institution{Qwant\\
  Georgia Tech \& UTC}
  \city{Paris}
  \country{France}
}

\author{Matyas Amrouche}
\email{m.amrouche@qwant.com}
\affiliation{%
  \institution{Qwant}
  \city{Paris}
  \country{France}
}

\author{Romain Deveaud}
\email{r.deveaud@qwant.com}
\affiliation{%
  \institution{Qwant}
  \city{Paris}
  \country{France}
}

\newcommand{\nb}[3]{{\colorbox{#2}{\bfseries\sffamily\scriptsize\textcolor{white}{#1}}}
{\textcolor{#2}{\sf\small\textit{#3}}}}
\newcommand{\rrdd}[1]{\nb{RD}{orange}{#1}}

\renewcommand{\shortauthors}{Fitte-Rey et al.}
\begin{abstract}
Building high-quality datasets and labeling query-document relevance are essential yet resource-intensive tasks, requiring detailed guidelines and substantial effort from human annotators. This paper explores the use of small, fine-tuned large language models (LLMs) to automate relevance assessment, with a focus on improving ranking models' performance by augmenting their training dataset. We fine-tuned small LLMs\footnote{A Language Model that is composed of 7B parameters is definitely not small, although likely two orders of magnitude smaller than current GPT models. We believe that the term "small LLM" can accurately describe such smaller models, in addition to being a nice oxymoron.} to enhance relevance assessments, thereby improving dataset creation quality for downstream ranking model training. Our experiments demonstrate that these fine-tuned small LLMs not only outperform certain closed-source models on our dataset but also lead to substantial improvements in ranking model performance. These results highlight the potential of leveraging small LLMs for efficient and scalable dataset augmentation, providing a practical solution for search engine optimization.
\end{abstract}

\maketitle

\section{Introduction}
In recent years, there has been tremendous interest in using Large Language Models (LLMs) to automatically provide relevance judgments for query-document pairs~\cite{abbasiantaeb2024can,thomas2024large,upadhyay2024llms}. Industry practitioners usually heavily rely on human annotations to create datasets for offline evaluation or training ranking models. Depending on the scale of such datasets, this process can be extremely costly, error-prone, and requires significant quality control~\cite{alonso2019}. The promise of relatively inexpensive relevance judgments, alongside initial results showing the capabilities of LLMs in replicating human annotations, has sparked strong interest from the information retrieval (IR) community.

However, a large portion of these experiments relies on closed-source LLMs accessed via APIs, which raises concerns regarding reproducibility -- how can we know which version of the LLM is responding at any given time? --, cost, and lack of control. Moreover, there have been concerns about using LLMs for relevance assessments~\cite{faggioli2023}, as well as doubts about their effectiveness in reproducing human judgments~\cite{clarke2024llmbasedrelevanceassessmentcant}.

In this paper, we take a slightly different approach and explore how small LLMs can be used to rescale datasets for training ranking models, such as dense encoders or traditional Learning to Rank models. Ranking models perform best at scale when presented with a large number of counter-examples of varying difficulty during training. These negative examples are often automatically generated by sampling from distributions of documents ordered by different ranking functions, each with varying effectiveness depending on the required difficulty of the negatives. In this context, false negatives may be introduced, which can harm the quality of the final ranking models.

We experiment with using small LLMs to adjust the relevance labels of automatically generated ranking datasets and evaluate this process on a re-ranking downstream task by fine-tuning a bi-encoder. This approach avoids the costs and lack of control associated with closed models by performing fine-tuning experiments on open-source LLMs, such as Llama 3.1, Gemma 2, and Qwen 2.5, for a Web search labeling task.

The contribution of this paper is twofold: 1) we explore how small LLMs can be effectively used for relevance annotation by fine-tuning them for a specific Web search task, providing better control over generations at a fraction of the cost, and 2) we apply these models to a re-ranking downstream task, where LLMs adjust the relevance of semi-automatically generated ranking datasets. Our results demonstrate that small fine-tuned models can compete with much larger closed models, such as GPT-4, in relevance annotation tasks. Evidence from our experiments also shows that automatically adjusting relevance labels with such models consistently improves the quality of ranking models.


 
\section{Related Work}

\subsection{Large language Models}

LLMs have revolutionized NLP by enabling advanced understanding and generation of human language across diverse applications. This paradigm shift began with models like BERT~\cite{alaparthi2020bidirectional}, which showcased the ability to discern complex contextual relationships between words and phrases through pretraining on vast amounts of unlabeled text data. Building on this foundation, T5~\cite{raffel2020exploring} introduced a unified text-to-text framework, enabling seamless adaptation to a wide range of NLP tasks, including direct relevance scoring for complex IR scenarios.

In the context of retrieval, LLMs have advanced the encoding of queries and documents by capturing deeper semantic meanings and relationships, surpassing traditional word-matching techniques. Models such as ColBERT~\cite{khattab2020colbert}, an optimized version of BERT, and E5~\cite{wang2022text}, which leverages contrastive learning, have set benchmarks for embedding-based approaches by significantly improving retrieval performance. More recently, decoder-only architectures, such as those in the GPT~\cite{radford2018improving} family, have gained prominence. These include both proprietary models, like OpenAI’s GPT series, and open-source alternatives such as Llama~\cite{touvron2023llama}  and Gemma~\cite{team2024gemma}.

While scaling up LLMs has been a dominant trend for maximizing performance, efforts to develop efficient, smaller models have gained traction. For example, Gemma 2~\cite{team2024gemma2} provides models ranging from 2 billion to 27 billion parameters, achieving state-of-the-art results through innovations like local-global and group-query attention mechanisms, which enhance scalability and efficiency. Given our resource constraints and the need for fast inference in production-ready applications, smaller, resource-efficient models such as Gemma 2 are particularly relevant, as they reduce dependency on extensive GPU resources while maintaining competitive performance.

\subsection{Ranking using LLM}
An emerging trend in leveraging LLMs for ranking tasks involves directly providing a query and a set of documents, asking the models to rank the documents based on relevance. While this approach suggests the possibility of replacing traditional relevance assessment methods, the results have often fallen short of expectations. Despite their ability to capture nuanced semantic relationships, LLMs frequently underperform compared to state-of-the-art deep learning models, particularly in terms of ranking accuracy and scalability. Several studies~\cite{ma2023zero,nogueira2020document} have attempted to fine-tune LLMs using ranking loss functions~\cite{zhuang2023rankt5}, achieving promising results, yet these models still struggle with issues such as handling large-scale datasets and producing consistent rankings.

Recent research has explored novel approaches to improve LLM-based ranking. For instance,~\cite{pradeep2023rankzephyr} investigates listwise reranking by distilling knowledge from proprietary models like GPT-4 to train a new ranker. However, this method often suffers from limitations such as omitting certain documents or generating repetitive rankings, which can degrade performance. Similarly, ~\cite{qin2023large} explores pairwise ranking using a sliding window approach to rank the top-k documents. While effective in certain scenarios, this method requires multiple passes over the dataset to generate a final ranking, making it computationally expensive and impractical for real-world applications involving large datasets.

\subsection{Relevance assessment using LLM}
Relevance assessment is a cornerstone of IR, focusing on determining how well a document satisfies a user's query. Traditional approaches relied on human annotations, which, while accurate, were resource-intensive.

LLMs have emerged as a transformative tool for relevance assessment, offering the potential to replicate or enhance human annotations at scale. Recent studies~\cite{abbasiantaeb2024can,macavaney2023one,thomas2024large,upadhyay2024llms} demonstrate that LLMs can approximate user judgments, fill annotation gaps, and align labels with crowd-sourced annotations. Single-shot methods~\cite{macavaney2023one} generate relevance labels efficiently but often lack depth, while multi-shot approaches, offering richer context, improve performance but are computationally more demanding. 

Despite these advancements, challenges persist. Many studies rely on closed-source models prone to API updates, while the exploration of smaller, efficient LLMs remains limited. Additionally, most research focuses narrowly on replicating human labels rather than leveraging LLMs for tasks like embedding training, ranking optimization, or dataset denoising. Our work aims to extend these findings by exploring the practical applications of fine-tuned, smaller LLMs for relevance assessment in real-world scenarios, ensuring scalability and accuracy while addressing data quality challenges.

While \cite{ma2024fine} explores label creation through large-scale fine-tuning on extensive datasets, our work diverges by enhancing the basic query–document comparison with additional prompt attributes —such as query intent and document metadata. By fine-tuning small LLMs on a carefully curated set of only 1,000 examples, we offer a resource-efficient method that still produces high-quality relevance labels to augment downstream industry-grade ranking tasks.

\section{Methodology}

The goal of this paper is twofold.
First, we focus on the capabilities of small LLMs (around 7–9B parameters) to accurately reproduce human-grade relevance judgments.
Then, we investigate whether these models can be applied on dataset improvement tasks and the effect this may have on downstream tasks such as top-$k$ re-ranking of Web search.
Specifically, we aim to answer the following research questions:
\begin{description}
\item {\bf RQ1}: Are small LLMs capable of replicating human labels?
\item {\bf RQ2}: Does fine-tuning increase the quality of relevance labels for Web search data?
\item {\bf RQ3}: Can LLM labeling improve the effectiveness of a Web search dense re-ranker?
\end{description}


The following sections detail our approaches for implementing query-document relevance labeling and fine-tuning the small LLMs.
They also detail our strategies for improving the quality of datasets that would be used to fine-tune a bi-encoder for top-$k$ re-ranking.

\subsection{Choice of LLM}
 Achieving scalability that aligns with the resource constraints of our production environment is a primary objective, alongside maintaining result quality. Therefore, leveraging LLMs was impractical to label millions of query-document pairs. Consequently, we focused on smaller LLMs, specifically those with 7 to 9 billion parameters, which offer a resource-efficient alternative for generating high-quality annotations, striking a balance between scalability and performance. 

Furthermore, while closed-source LLMs (such as GPT or Anthropic ones) are currently the most efficient ~\cite{huggingface2024} ~\cite{paperswithcode2024} and outperform other models, they present several challenges. Relying on third-party APIs introduces operational risks—services can experience downtime, models can be deprecated, or pricing and access conditions can change. For a search engine that integrates these models in its ranking pipeline, such disruptions could severely impact the system's functionality. Additionally, there might be concern about data privacy, as any information shared with these LLMs through their APIs might become accessible to the providing companies and can potentially be leaked by the LLM. This means private datasets used for training or labeling could potentially be exposed.



For all these reasons, this study focuses on small open-source LLMs. GPT-3.5 and GPT-4 serve as baselines, while models such as Llama 3.1 8B, Gemma 2 9B, and Qwen 2.5 7B are tested for comparison. These three models are also used for the fine-tuning part.

For the downstream re-ranking phase, the focus narrows to Llama 3.1 8B and Gemma 2 9B, as these models consistently delivered the best results across tasks.

All models have various model cards and some differences in terms of training techniques, therefore, the format (FP16, BF16) and other mechanisms (flash-attention...) are written on the results of each experiment.

\subsection{Prompting} \label{prompting}
This part outlines the methodology for addressing \textbf{RQ1}. The objective is to assess the ability of small LLMs to replicate human labels and to analyze the limits of the information that can be provided to improve replication.

Our prompting approach is based on~\cite{thomas2024large}. They conducted an extensive study on prompt variations, though their focus was on Text REtrieval Conference (TREC)\cite{chowdhury2007trec} evaluation, primarily examining query and passage interactions.

\begin{figure}[h] 
    \caption{Structure of the prompt used in our experiments, with bracketed text indicating placeholder inputs} 
    \includegraphics[width=\linewidth]{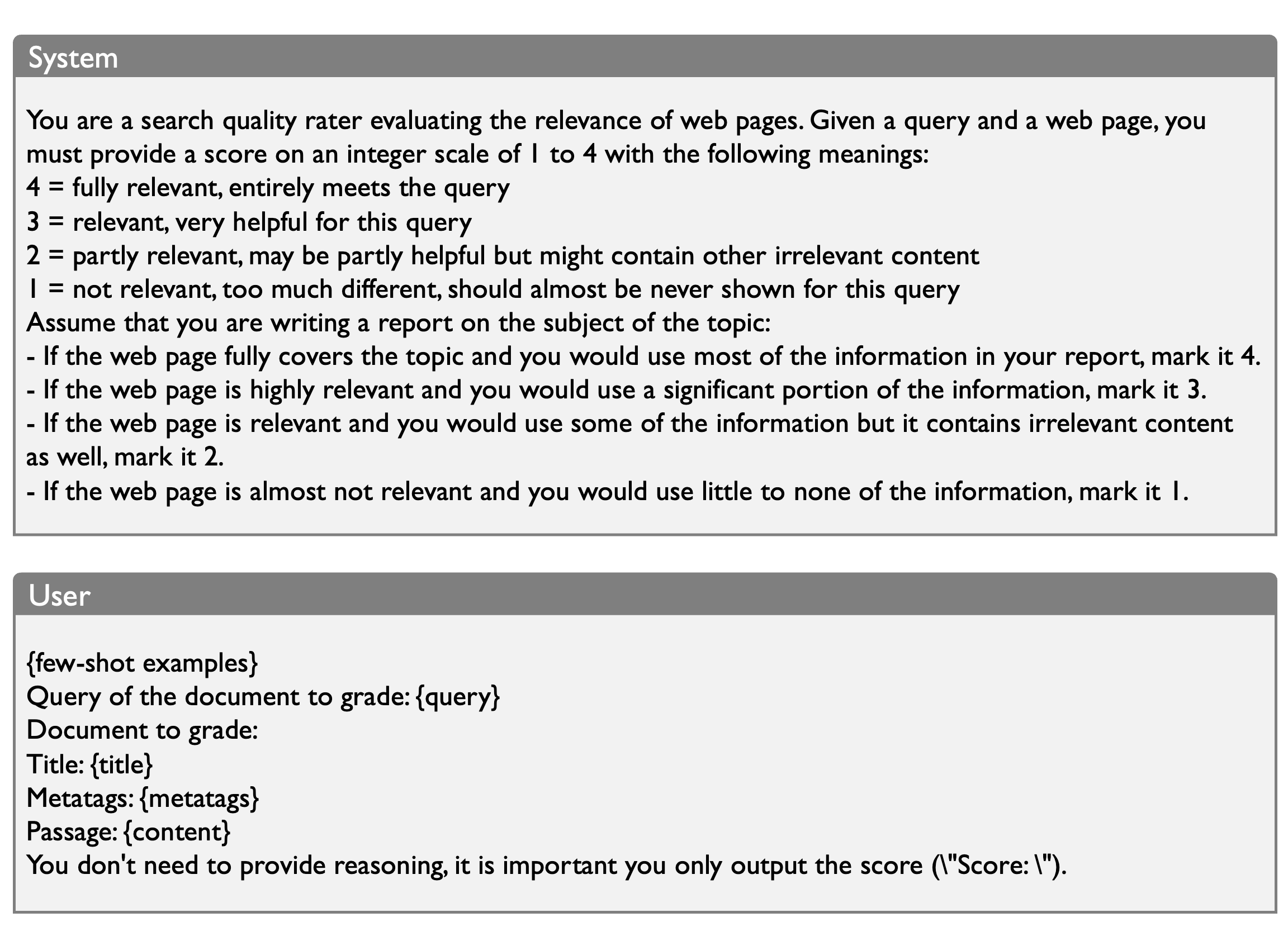} 
    \label{fig:prompt} 
\end{figure}

Our prompt, Figure \ref{fig:prompt}, is divided into four distinct parts. This prompt represents the final version of all the different tries, however, differences in wording or other small changes will not be detailed in this paper.

\subsubsection{System prompt / Role}
In most LLMs, this first part is known as the system prompt. It defines the role and provides instructions to the LLM. This section is largely similar to what’s described in \cite{thomas2024large} and aligns with the TREC assessor instructions. It begins by defining the LLM's role, a common prompting method to give context and a clear mission to the model. It then introduces a scale from 1 to 4, explaining the correspondence of each grade. The scale is further clarified with different wording to ensure the model fully understands the distinctions among the four grades, and to avoid any other grade being applied. Since we use small LLMs, repetition in the prompt can reinforce the requirements (we observed issues like grades outside the 1-to-4 range when repetition was lacking).

\subsubsection{Few-shot vs Zero-shot}
Depending on the type of experiment we are doing, before providing the query and the document to grade, we can add some examples to make some few-shot learning. The template is not different from what we will present for the document. The main difference is that the query and document elements are numbered so that the LLMs can easily determine which title and passage are associated. In most models’ cards, these examples are put in the user prompt but it can be in some occasions concatenated to the system prompt presented earlier.

\subsubsection{Document presentation}
Our document representation differs from approaches like ~\cite{macavaney2023one, thomas2024large} that follow strict TREC formatting. In our case, a document is more than just a title and passage—it includes the URL, extracted HTML elements (title, content, meta tags description...), and multiple metadata components.

The prompt starts with the query. Then, we have developed multiple options to enrich the prompt. While an initial prompt would be based on the first 250 characters of the document content and the title, we have developed additional strategies to provide the model with a more comprehensive document understanding.

We prompt the LLM with several document fields: the title and the most informative parts of the textual content, the {\tt meta} description tags when available, and the URL.
This allows the LLMs to take advantage of their background knowledge and recognize if the URL belongs to a trusted, well-known website.
We also provide the LLM with the output of an in-house query intent classification model that provides probability estimates over three well-known intents: navigational, informational, and transactional.

This can help the LLM understand the context and purpose of the query. For instance, if the user is researching a historical event, the model could prioritize Wikipedia or news sites as more relevant sources.

\subsubsection{Final instruction and output}
Finally, this section allows for modifying the objective of the gradation, clarifying the expected result. The chain-of-thought (COT) approach has become popular and can enhance the output by guiding the model to provide answers through step-by-step reasoning. This approach is particularly useful for complex prompts and challenging problems (the O1 model, for instance, is based on this step-by-step reasoning). This section also restates the goal and specifies the expected output format. We chose to avoid complicating it with a JSON output, instead expecting the word "Score" followed by the label. We also attempted an aspect-based approach, which required multiple grades on different comparison elements.

Our prompt might not be the most optimal, but it performed best for our needs and ensures experimental comparability across different LLMs. The objective of this paper is not to try many prompts but rather provide ones that are effective on many LLMs with some options to increase the results. A detailed evaluation in the experiment part will show the gain and the loss with the options, methods (COT), and the results are different from other papers as smaller models do not possess the same capability of extracting and analyzing information.

\subsection{Fine-Tuning}
Fine-tuning has been used in many cases to enhance the model on a specific task. It usually comes with decreased performance in other tasks, but depending on the objectives, this may not be an issue. If maintaining performance on other tasks matters, fine-tuning should be done carefully to avoid issues such as catastrophic forgetting.

In our case, we decided to fine-tune 3 different models to enhance the labeling. This part details the methodology for addressing \textbf{RQ2}. The objective is to align the predictions of relevance with our dataset, which contains human-defined golden labels. This approach aims to better differentiate documents that fully meet the query when compared to the vanilla version.

When it comes to fine-tuning, a key factor is the dataset. To create this dataset, we used findings from \cite{zhou2024lima}, which demonstrated that 1,000 carefully curated prompts and responses are more effective than a larger dataset that may contain noise or contradictions. We manually reviewed 1,250 prompts (queries and documents) to obtain 1,000 prompts for training and 250 for validation during training. All relevance classes are evenly distributed (25\% each), as experiments showed that an imbalance leads the model to predominantly predict the majority class \cite{abbasiantaeb2024can}.


We performed LoRA~\cite{hu2021lora} fine-tuning, as full fine-tuning LLMs is resource-intensive. The results section will present only the outcomes obtained from the best combination of parameters.

Fine-tuning was performed on A4500 GPUs in brain float-16 format, leveraging flash-attention~\cite{dao2023flashattention} to enhance training efficiency. Inference was carried out in float-16 on 8 V100 GPUs, showcasing the stability of the results.

To comprehensively evaluate the effectiveness of fine-tuning beyond the validation dataset, we apply the same dataset used in the prompting section to assess improvements in model performance during its fine-tuning. Only the results from the best fine-tuning iteration for each model are presented.

\subsection{LLM relevance assessment evaluation}
There are various ways to evaluate labels. One approach might be to generate labels similar to those considered the best, typically created by human annotators, a pointwise comparison. However, we must consider whether the relevance value itself is crucial or if the relative order (listwise comparison) between documents matters more. The relative order between documents can be particularly interesting to evaluate because, if a model has a bias (as shown for GPT ~\cite{abbasiantaeb2024can}), it will perform poorly in direct comparison. However, it may still achieve similar results if the relative scores are consistent, allowing any ranking model to learn similarly. This creates an ordering that we could evaluate by query. Alternatively, we could evaluate by comparing pairs of documents based on their scores, using a method resembling pairwise ranking ~\cite{qin2023large}, but we did not explore it.

Regarding the agreement between original labels and LLM generated ones, the comparison is straightforward. In this paper, we reported two different metrics to assess label differences. First, we used Cohen's Kappa metric, which measures agreement: a value of 1 indicates that the generated labels perfectly match the original labels, meaning our model aligns well with the ground truth, while -1 indicates complete disagreement, suggesting poor alignment. This metric ranges between these values. Second, we used the mean absolute error (MAE), which reflects the error made by the LLM across all labels. MAE can also be differentiated by queries for analysis. We use a binary version, where 1 indicates disagreement between original and LLM labels and 0 indicates agreement.

As previously discussed, assessing agreement based solely on absolute values can be limiting, as it fails to capture the importance of relative document rankings. To address this, we focused on methods and metrics that evaluate agreement based on the relative order of documents. For instance, if the original labels for three documents are 2, 3, and 4, a model that predicts 1, 2, and 3 should not be penalized, as the relative order remains intact. This correctly reflects that the last document is more relevant than the second. To measure relative agreement, we utilized Kendall’s Tau, a metric designed to evaluate the alignment of rankings. Specifically, Kendall’s Tau quantifies the agreement between human-labeled and LLM-generated rankings for documents within a single search engine results page (SERP). By capturing the concordance and discordance of pairs, Kendall’s Tau effectively reflects how well the two rankings align. The metric ranges from 1 (complete agreement) to -1 (complete disagreement), making it particularly useful for comparing ordered lists in relevance tasks.


\subsection{Bi-encoder fine-tuning on relevance dataset rescaled by small LLMs}

When fine-tuning bi-encoder models for downstream ranking tasks, one key factor of ranking effectiveness is the quality of negative examples in the training data \cite{robinson2020contrastive}. More specifically, regarding hard negatives, negatives which are difficult to distinguish from positive examples.
Indeed, false hard negatives can harm training by providing contradictory signals.

\begin{figure}[h] 
    \caption{Schema representing the rescaling process of the dataset by dividing the hard negatives into soft-like negatives, hard negatives, and false hard negatives sub-buckets.} 
    \includegraphics[width=\linewidth]{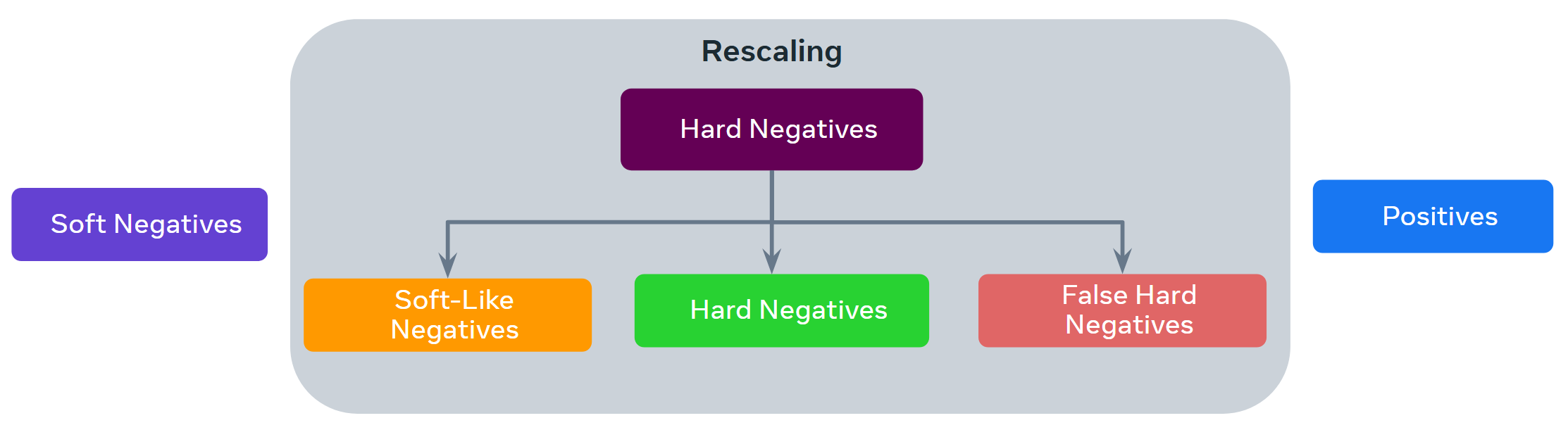} 
    \label{fig:rescl} 
\end{figure}

When inspecting our custom ranking datasets, we found that approximately 10\% of the hard negative samples were false hard negatives, thus significantly degrading the model's ability to delimit a reliable frontier between positive and negative examples. This issue arises because hard negatives sampling is inherently challenging, as it is difficult to ensure that all negative samples are truly irrelevant or appropriately hard negatives. We hypothesized that leveraging LLMs specialized in relevance judgments could help assess the quality of hard negative samples and adjust their labels accordingly. By correcting these false labels, we can enhance the dataset's quality, ensuring better ranking performance.

We considered a proprietary query-document ranking dataset, made of query and document content text pairs, that includes for each query soft negative, hard negative, and positive examples.


The process involved rescaling (Figure \ref{fig:rescl}) the dataset hard negatives' relevance labels with our relevance-specialized small LLMs into three distinct relevance buckets: \begin{itemize} \item \textbf{Soft-like negatives}, examples flagged as SLN (label = 1) by the LLM, \item \textbf{True hard negatives}, examples flagged as HN (label = 2) by the LLM, \item \textbf{False hard negatives}, examples flagged as FHN (label = 3 or 4) by the LLM. \end{itemize}

The bi-encoder's backbone model used in this task is a pre-trained E5 multilingual~\cite{wang2024multilingual}. Pre-trained E5 models were then fine-tuned on both the original and the rescaled datasets, to be finally evaluated on the same original evaluation dataset (the dataset with the non-rescaled relevance). This experiment demonstrated that training on a rescaled dataset translated to better ranking performance when applied back on the original dataset, ultimately addressing \textbf{RQ3}.

To assess the impact of our fine-tuned small LLMs relevance rescaling on the downstream ranking quality, we used two evaluation metrics: \begin{itemize} \item \textbf{NDCG@k}, which measures the overall SERP quality. \item \textbf{MRR}, which assesses whether the first relevant document appears in the top positions. \end{itemize}

\section{Experiments}
We defined several experiments to evaluate our methods. Firstly, we will evaluate the prompting and address \textbf{RQ1}. To do this, we created a dataset of one thousand queries from our real-world data, which we cleaned to ensure meaningful results. For instance, we limited the number of negative elements in each SERP to avoid inflating metrics, as LLMs can easily detect these in relation to a query. As a result, the dataset contains an average of 5.6 documents per query, and the distribution of each relevance class is more balanced, as shown in table \ref{tab:relevance-distribution}. The possible classes range from 1 to 4, where 1 represents "not relevant," 2 represents "partly relevant," 3 represents "relevant," and 4 represents "fully relevant". This setup is more challenging than the typical TREC evaluation, which only uses two classes. We will evaluate different prompting strategies with various models and use closed-source models as baselines (GPT-3.5 and GPT-4, with GPT-4 providing better predictions than GPT-4o).

Next, we will present the variations in performance achieved through fine-tuning. We used the best prompting strategy in terms of both performance and efficiency, ensuring that the results specifically improve with this approach. This part addresses \textbf{RQ2}.

Finally, we will demonstrate how the labels can be used to improve the relevance of a real-world dataset for training a dense ranker. This will show that reducing noise and incorporating more detailed relevance levels enhance performance. The evaluation of the ranker addresses \textbf{RQ3}.

\begin{table}[t]
  \caption{Distribution of relevance across classes in the dataset used to evaluate prompting methods (1: Not Relevant, 2: Partly Relevant, 3: Relevant, 4: Fully Relevant)}
  \label{tab:relevance-distribution}
  \centering
  \begin{tabular}{lcccc}
    \toprule
    \textbf{Class} & \textbf{1} & \textbf{2} & \textbf{3} & \textbf{4} \\
    \midrule
    \textbf{Frequency (\%)} & 32.85 & 28.5 & 17.2 & 21.45 \\
    \textbf{Count}          & 1854  & 1608  & 971  & 1211  \\
    \bottomrule
  \end{tabular}
\end{table}

\subsection{Prompting}
Table \ref{tab:prompting-performance} summarizes the comparison between LLM-generated labels and the human ones from our dataset, focusing on the labels themselves and their relative order within a SERP, evaluated using the three metrics detailed above. The dataset consists of approximately 5,600 query-document pairs. The rows present the results for each model associated with a specific prompting strategy, along with various options (each checked box indicates an activated option, and these options can be cumulative). For instance, if both U and I are checked, it indicates that both the intention and the URL are included in the prompt. The best score for each metric across the models is highlighted in bold.

Overall, we observe that performance varies primarily by model. As expected, GPT-4 achieves the best results with a 10-example few-shot approach. This outcome is logical, as GPT-4 is theoretically the largest model released, enabling it to leverage a vast amount of information and establish a scale based on that knowledge. Interestingly, all models perform better in terms of relative ranking than in absolute label agreement. The maximum $\kappa$ score is only 0.34, whereas $\tau$ reaches up to 0.67. This suggests that even when the labels do not perfectly match, their relative order within a query remains consistent. This phenomenon has been observed in previous studies, indicating that models may exhibit a bias toward certain labels, predicting coherent relative rankings even when exact labels are incorrect.

Our experiments indicate that GPT-3.5 shows lower performance compared to newer, smaller LLMs under specific prompting strategies. Models such as LLama, Gemma, and Qwen outperform GPT-3.5 with certain strategies across all three metrics. This highlights that recent advancements in pre-training (e.g., distillation \cite{hinton2015distilling}) and post-training techniques (e.g., RLHF \cite{ouyang2022training}) using enhanced datasets enable more efficient knowledge extraction without requiring a larger number of parameters.

A general observation is that even with a four-level relevance scale, the LLMs demonstrate a solid understanding of labeling and relevance variation. Although not tested here, we are confident that smaller LLMs would perform well on academic datasets like TREC-Robust, which include only 2–3 classes.

\subsubsection{Prompts options}

\begin{table}[t]
  \caption{Performance evaluation of prompting strategies (U=URL, I=Intention, COT=Chain of Thought, F4/F8=Few-shot with 4/8 documents) on human-annotated query-document pairs, measured through document label accuracy (MAE, $\kappa$) and relative order ($\tau$). Bold values indicate best performance}
  \label{tab:prompting-performance}
  \centering
  \small 
  \begin{tabular}{|l|ccccc|cc|c|}
    \toprule
    \textbf{Model} & \textbf{U} & \textbf{I} & \textbf{COT} & \textbf{F4} & \textbf{F8} & \multicolumn{2}{c|}{\textbf{Doc scores}} & \textbf{Doc order} \\
    
                   & & & & & &  \textbf{MAE} & \textbf{$\kappa$} & \textbf{$\tau$} \\
    \midrule
    GPT-3.5     & &  &  &  &  &  0.63 & 0.16 & 0.40 \\
    GPT-3.5      & \checkmark & \checkmark  &  & &   & 0.60 & 0.19 & 0.48 \\
    GPT-3.5     &  &  &  & & \checkmark    & 0.65 & 0.15 & 0.47 \\
    GPT-3.5      &  &  & \checkmark  &   &  & 0.57 & 0.22 & 0.45 \\
    \midrule
    GPT-4     & &  &  &  &  &  \textbf{0.47} & 0.32 & 0.59 \\
    GPT-4      & \checkmark & \checkmark  &   &&  &   0.48 & \textbf{0.34} & 0.63 \\
    GPT-4     &  &  &  &  & \checkmark  &  0.48 & \textbf{0.34} & \textbf{0.67} \\
    GPT-4      &  &  & \checkmark  & &  &   0.49 & 0.33 & 0.63 \\
        \midrule
    Llama 3.1 8B &  &  &  &  &   &  0.57 & 0.22 & 0.46 \\
    Llama 3.1 8B &  &\checkmark  &  &   & &   0.58 & 0.21 & 0.47 \\
    Llama 3.1 8B & \checkmark &  &  & &  &   0.56 & 0.23 & 0.49 \\
    Llama 3.1 8B & \checkmark & \checkmark &    & &  & 0.57 & 0.22 & 0.49 \\
    Llama 3.1 8B & &  &  & \checkmark &   &  0.56 & 0.23 & 0.50 \\
    Llama 3.1 8B & &  &  &  & \checkmark  &  0.57 & 0.21 & 0.46 \\
    Llama 3.1 8B &  &  &  \checkmark&  & &   0.52 & 0.25 & 0.50 \\
        \midrule

       Gemma 2 9B&  &  &  &  &  &   0.56 & 0.25 & 0.54 \\
       Gemma 2 9B & &  \checkmark &  &  & &   0.55 & 0.26 & 0.57 \\
     Gemma 2 9B&  \checkmark & &  &  & &   0.54 & 0.27 & 0.58 \\
     Gemma 2 9B  & \checkmark &  \checkmark &  & &  &   0.53 & 0.29 & 0.62\\
     Gemma 2 9B &  &  &  &  \checkmark & &   0.54 & 0.27 & 0.64 \\
     Gemma 2 9B& &  &  &  &   \checkmark  &  0.54 & 0.27 & 0.64 \\
    Gemma 2 9B & &  &  \checkmark &  & &   0.53 & 0.29 & 0.58 \\
        \midrule
            Qwen 2.5 7B &  &  &  &  &  &   0.56 & 0.23 & 0.50 \\
            Qwen 2.5 7B  & &  \checkmark &  &  & &   0.55 & 0.25 & 0.52 \\
     Qwen 2.5 7B &  \checkmark & &  &   & &   0.56 & 0.23 & 0.51 \\
      Qwen 2.5 7B& \checkmark &  \checkmark &  &  & &   0.55 & 0.26 & 0.54 \\
           Qwen 2.5 7B  &  &  &  &  \checkmark &   &  0.55 & 0.25 & 0.53 \\
          Qwen 2.5 7B & &  &  &  &   \checkmark &  0.55 & 0.26 & 0.54 \\
          Qwen 2.5 7B & &  &  \checkmark &   & &   0.55 & 0.25 & 0.53 \\
    \bottomrule
  \end{tabular}
\end{table}

We have explored various prompting strategies and will now summarize our findings, beginning with the negative results.

The first issue involves Aspects. We had high expectations for aspects, which were introduced in ~\cite{thomas2024large} and involve grading relevance across multiple aspects before providing a final score. While this approach yields significant improvements in ~\cite{thomas2024large}, our results are more mixed. The main problem, particularly with Llama, Gemma, and Qwen, is twofold: either the expected format is incorrect (with one aspect or the global grade missing), or the results are heavily biased, with the final score being consistently the same. Consequently, our experiments lead us to dismiss the use of aspects, as they prove too challenging for smaller LLMs to interpret and utilize effectively for reasoning.

Secondly, providing the URL alone results in mixed outcomes. For Gemma, it improves predictions, but this is not the case for Qwen or Llama. This is unexpected, as the URL should provide additional context about the document, making it easier to determine if it matches the query—especially in cases where the content is ambiguous (e.g., e-commerce sites). Upon examining the predictions, we observe that when the URL is provided alone, Llama and Qwen often attempt to match it exactly to the query. As a result, if the URL does not precisely match the main word or site in the query, it frequently leads to very low predictions.

In contrast, the intent feature shows clear positive impacts. Adding intent consistently improves model performance. From Table \ref{tab:prompting-performance}, we see that all metrics improve whenever intent is included. We present the intent by providing a brief definition or example of use cases. This improvement is particularly noticeable for navigational queries, where the LLM becomes more stringent in identifying the correct website (e.g., penalizing AccuWeather if the query is Meteo France).

While the URL alone does not always yield improvements, combining intent with the URL significantly enhances the prompting. This combination greatly improves the relative ranking of documents (e.g., +14.8\% on Gemma, +20\% on GPT-3.5, +8\% on Qwen, and 6.78\% on GPT-4), demonstrating that the URL becomes useful when paired with sufficient context. The URL is particularly helpful for navigational queries, where it makes it easier to identify the correct website, and for informational queries, where it helps identify sources like Wikipedia or news sites.

Using few-shot examples also improves predictions. All models benefit, but the most significant gains are observed in larger models (e.g., GPT-4 +0.08 and Gemma 2 +0.1 for $\tau$). Smaller models show less reliable improvements (e.g., Llama benefits from a 4-example few-shot approach but not an 8-example one). Smaller models are more prone to biases introduced by the examples. For instance, when we use only one or two examples, the predicted relevance classes often match those examples. To address this, we provide a number of examples that match the number of classes, ensuring representation for each class. Even with this precaution, confusion matrix analyses reveal biases introduced by the order of the examples. As a result, few-shot prompting can be a useful option but should be applied carefully, especially with smaller models. It is more effective with larger models or Gemma, where examples must be chosen meticulously. Moreover, this approach increases prompt length, thereby reducing LLM throughput and increasing inference cost.

Finally, COT consistently improves predictions. Allowing the model to articulate its reasoning for a grade leads to more precise outcomes. This has been demonstrated in prior studies \cite{wei2022chain}, and we confirm its utility in the context of labeling. The only drawback of COT is reduced throughput: generating hundreds of tokens for reasoning takes significantly longer than predicting just a few tokens, which becomes a major limitation when processing millions of predictions.

Overall, in relation to \textbf{RQ1}, we conclude that smaller LLMs can replicate human labels most of the time, particularly in terms of relative ranking rather than exact label values. While the results are not perfect—especially when compared to the GPT-4 baseline—fine-tuning has the potential to yield significant improvements.

\subsubsection{Prompting influence on LLM throughput for Gemma 2 9B}
One of the major issues that led us to use smaller LLMs is the need to optimize throughput. Given the resources and time constraints for end-to-end project iterations, minimizing latencies and maximizing throughput are critical factors, highlighting the necessity of considering smaller LLMs. However, the size of the LLM is not the only factor influencing throughput; both the prompt length and the length of the expected prediction also play significant roles.

Our basic prompting approach only asks for the output to be in the format 'Score: value,' which ensures fast inference for each prompt. As previously discussed, the COT method improves performance but also increases the number of tokens to predict—from 2 or 3 tokens to hundreds or even thousands. Consequently, it becomes impractical to use COT in a cost-efficient solution. Table \ref{tab:prompting-throughput} presents the average throughput for 5,000 prompts from our experiments with the Gemma model, conducted on 8 V100 16 GB GPUs using the VLLM package with FP16 inference and pipeline parallelism. The results show that COT significantly degrades throughput, reducing it from 45.9 to 6.27 prompts/sec, making it unsuitable for scenarios involving the annotation of millions of pairs for training dataset creation. This reduction is primarily due to the increased prompt length.

Few-shot prompting also affects throughput as the prompt length increases by a factor of 2.5 or 4.5, depending on the number of examples included. Although the impact is less severe than with COT (since processing for few-shot prompting involves only one pass rather than iterative generation), it still renders the approach impractical for production use. For example, throughput is more than halved (to below 50\%) when using 4 examples (F4), which is not feasible for a search engine.

This analysis underscores why, for fine-tuning and subsequent steps, we prioritize a prompt design that excludes COT and few-shot examples to maximize throughput. Instead, we focus on the best prompting method without these elements, incorporating intent and URL into the basic prompt for optimal performance.

We observe similar results with Llama and Qwen, with variations primarily driven by the specific size of the model.

\begin{table}[t]
  \caption{Throughput of different prompting methods of Gemma 2 9B using the prompt presented in \ref{prompting} (Score=Basic prompt, COT=Chain of thought, F4/F8=Few-shot with 4/8 documents)}
  \label{tab:prompting-throughput}
  \centering
  \begin{tabular}{lccccc}
    \toprule
    \textbf{Prompting Method} & \textbf{Score} & \textbf{COT} & \textbf{F4} & \textbf{F8}  \\
    \midrule
    \textbf{Prompt length}& 393 & 393 & 1031 & 1701 \\
    \textbf{Predicted token length }& 3 & 250 & 3 & 3 \\
    \textbf{Throughput (prompts/sec)}& 45.9 & 6.27 & 17.93 & 11.03 \\
    \bottomrule
  \end{tabular}
\end{table}

\subsection{Fine-Tuning}
This section addresses \textbf{RQ2}. To explore this question, we used the best prompting strategy described earlier while ensuring fast inference. This prompt was applied to a human-labeled dataset, and fine-tuning was conducted on three models: Llama, Gemma, and Qwen.

\begin{table}[t]
  \caption{Comparison of model performance between vanilla and fine-tuned configurations on human-annotated dataset, measured through document label accuracy (MAE, $\kappa$) and relative ranking order ($\tau$). Each vanilla configuration represents the best performing prompting strategy from Table \ref{tab:prompting-performance}}
  \label{tab:model-performance}
  \centering
  \small 
  \begin{tabular}{lcccc}
    \toprule
    \textbf{Model} & \textbf{Configuration} & \multicolumn{2}{c}{\textbf{Doc Scores}} & \textbf{Doc Order} \\
                   &                         & \textbf{MAE} & \textbf{$\kappa$} & \textbf{$\tau$} \\
    \midrule
    \textbf{GPT 4} & Vanilla   & \textbf{0.48} & 0.34 & 0.67 \\
    \midrule
    \textbf{Llama 3.1 8B} & Vanilla   & 0.57 & 0.22 & 0.49 \\
                          & Fine-tuned & 0.50 & 0.33 & 0.65 \\
    \midrule
    \textbf{Gemma 2 9B}   & Vanilla   & 0.53 & 0.29 & 0.62 \\
                          & Fine-tuned & 0.49 & \textbf{0.35} & \textbf{0.71} \\
    \midrule
    \textbf{Qwen 2.5 7B}  & Vanilla   & 0.55 & 0.26 & 0.54  \\
                          & Fine-tuned & 0.55 & 0.26 & 0.54 \\
    \bottomrule
  \end{tabular}
\end{table} 

Table \ref{tab:model-performance} summarizes the results obtained on the same dataset as used for prompting. The first row for each model restates the performance of the vanilla model with the same prompting strategy. The subsequent rows present the results achieved with the best fine-tuning epochs and parameters.

As anticipated, fine-tuned models demonstrate significant performance improvements. Gemma, which already performed well without fine-tuning, becomes the best model, achieving a $\tau$ of 0.71. This indicates that the relative order predicted by the LLM aligns closely with the human rankings. Furthermore, it outperforms GPT-4, which previously held the top position with COT prompting, even though Gemma now achieves this without using few-shot examples or COT. This result highlights a model that is both highly effective and efficient. Llama also performs well, achieving a $\tau$ of 0.65 for document order. This represents a significant improvement over its vanilla version (+0.16). An important consideration is that Llama is lighter than Gemma, which results in higher throughput. Depending on the application, this trade-off between performance and efficiency may be negligible, as discussed in the next section.

Qwen, however, struggles with fine-tuning. It fails to learn clear distinctions between relevance classes, particularly for classes 2 and 3, which are the most difficult to differentiate due to their proximity in definition (partly relevant vs. relevant). Nonetheless, distinguishing these classes is crucial in practice.

\begin{figure}[h] 
    \caption{Evolution of classification metrics (Precision, Recall, F1-score, and Accuracy) across fine-tuning epochs for Gemma 2 9B model on validation data, shown separately for each relevance class} 
    \includegraphics[width=\linewidth]{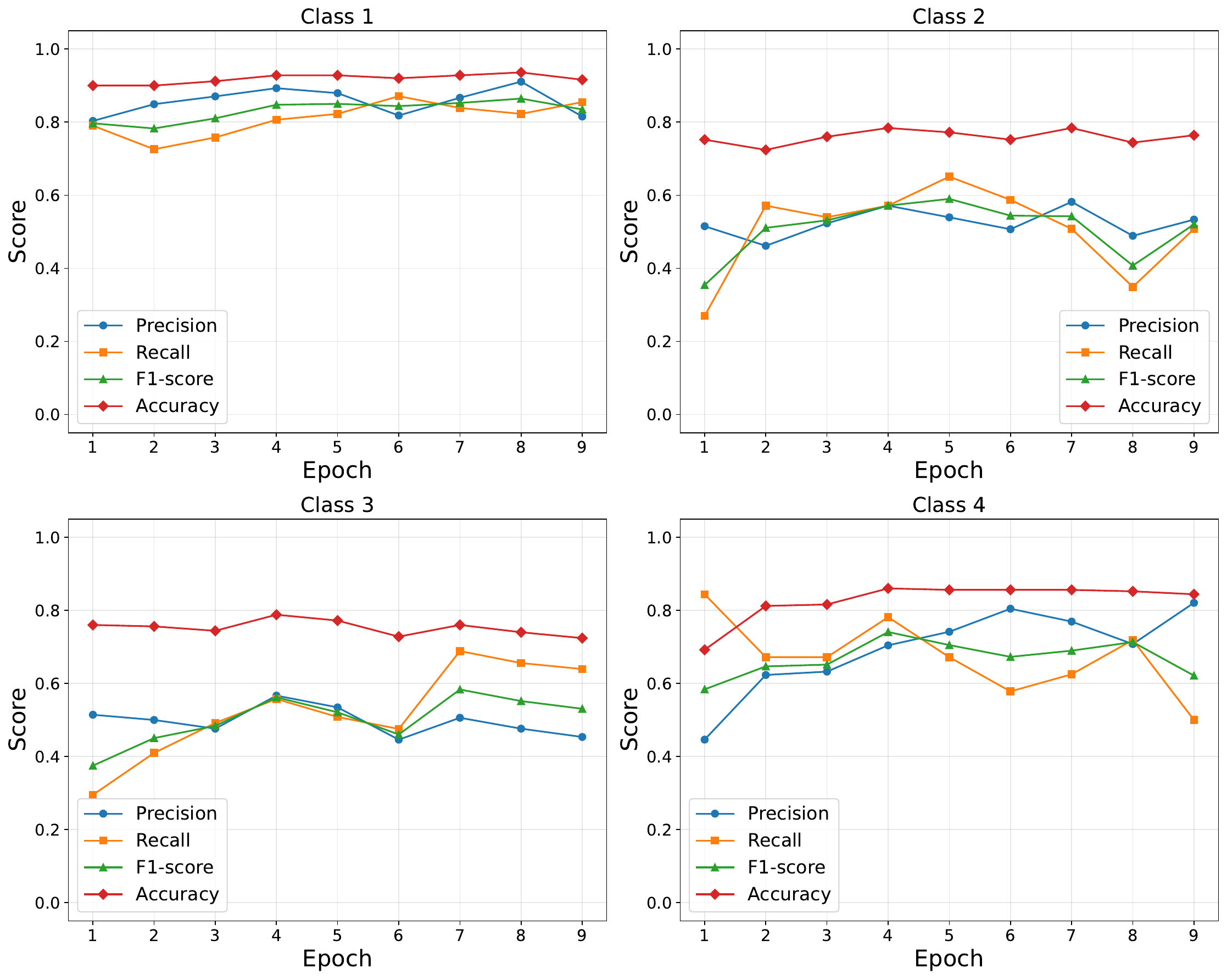} 
    \label{fig:gemmaTraining} 
\end{figure}

To better understand the learning process, Figure \ref{fig:gemmaTraining} illustrates the evolution of classification metrics during Gemma's fine-tuning on the validation dataset. Classes 1 and 4 are easier to predict as they represent the extremes, and prior research has shown that LLMs excel at distinguishing clearly relevant from irrelevant pairs. The intermediate classes (classes 2 and 3 which represent "partly relevant" and "relevant"), however, are more challenging. Initially, Gemma also struggles with these classes but successfully differentiates them by epochs 5 to 7. In contrast, Qwen continues to exhibit low precision and recall for the second class, with values remaining below 0.3.


One important remark from Table \ref{tab:model-performance} is that, while fine-tuning improves the results, the document score metrics remain relatively low. This indicates that the predicted scores are not exactly the same as the human labels, even though the ranking order is consistent. This discrepancy is not a critical issue, as previously shown in \cite{abbasiantaeb2024can}, but it does mean that mixing original labels with new ones should be avoided. For example, filling in gaps in a SERP where some documents are missing relevance scores will not perform as well as replacing all the labels. This is an important consideration when applying the labels to downstream tasks such as ranking. Therefore, in the next section, we apply our labeling method to all the documents in a SERP rather than just a subset of them.

\subsection{Ranking training}
In this part, we check if the relevance assessments made by vanilla and fine-tuned LLMs using our prompting strategy are transferable to a downstream ranking task. As we trained models capable of replicating human labels, we attempt to improve our ranking pipeline. The dataset used to fine-tune the E5 models consists of 18k training queries and 1k validation queries. Each query contains around 200 documents, scaled across 5 levels of relevance (soft negatives, hard negatives, and 3 levels of positives). About 70\% of these documents are hard negatives (approximately 2.5 million), and these are the ones we aim to rescale to new levels of relevance. These new levels are positioned between soft negatives and positives (we do not shift documents to a different category). This is important because, as explained earlier, the generated labels do not perfectly correspond to the original labels; only the relative order between them is of high quality. For the validation dataset, the labels are not rescaled to showcase the model's enhanced learning ability.

Since the training relies on query-document text pairs, we only use the title and content variables (Figure \ref{fig:prompt}) in the prompting for rescaling.

We run the experiment using several dimensions, employing Matryoshka loss during fine-tuning:

\begin{itemize}
\item 768 dimensions (by default, E5 hidden states dimensions)
\item 64 dimensions
\item 32 dimensions
\item 768 binary dimensions
\end{itemize}

We compute all the metrics for each query in the dataset and average them across all queries. Then, we compute the 95\% confidence intervals through 1000 bootstrap iterations \cite{hesterberg2011bootstrap} to show the statistically significant differences between two distributions.

\begin{table}[t]
  \caption{Percentage distribution of rescaled hard negatives (HN) to soft-like negatives (SLN), hard negatives (HN), or false hard negatives (FHN).}
  \label{tab:hn-percentage}
  \centering
  \begin{tabular}{l|c|c|c}
    \toprule
    \textbf{} & \textbf{HN $\rightarrow$ SLN} & \textbf{HN $\rightarrow$ HN} & \textbf{HN $\rightarrow$ FHN} \\
    \midrule
    \textbf{Vanilla LLMs} & 85\% & 8\% & 7\% \\
    \textbf{FT LLMs}      & 85\% & 5\% & 10\% \\
    \bottomrule
  \end{tabular}
\end{table}
  \vspace{-0.3cm}  



Our analysis, shown in Table \ref{tab:hn-percentage}, revealed that models predominantly assign a relevance score of 1 (85\% of cases), classifying documents as "soft-like negatives." Fine-tuned models (Llama, Gemma) identified 5\% true hard negatives (score 2) and 10\% false hard negatives (scores 3-4), aligning with our preliminary dataset review expectations.

The vanilla models showed different patterns, with 8\% true hard negatives and 7\% false hard negatives. However, manual validation revealed that only 6 out of 100 samples labeled as true hard negatives were actually correct, suggesting minimal potential for performance improvement when using these rescaled labels.


Table \ref{tab:ndcg-performance} summarizes the NDCG and MRR metrics on the validation dataset for models trained on the vanilla dataset and those rescaled by Llama or Gemma (both fine-tuned and non-fine-tuned).

Let us first discuss the NDCG results. An increase in NDCG is observed for all models trained on the rescaled dataset, consistently outperforming the baseline. As expected, fine-tuned models always surpass their vanilla versions.

The improvement in NDCG is more pronounced at NDCG@5 and NDCG@10 than at NDCG@30. This can be explained by the concentration of relevant documents near the top ranks: while the ordering is not perfect, positive documents are predominantly grouped at the start. Beyond a certain point, all remaining documents are negatives, which diminishes the impact of ordering and leads to a leveling effect in relevance.

The NDCG increase is consistent across metrics, with an approximate gain of 0.008 between the baseline and the best-performing Gemma fine-tuned model, with slightly lower gains for NDCG@30. This indicates that even models not exclusively focused on rescaling positive documents benefit from redistributing this large set of hard negatives during the training phase. Fine-tuned models, which focus on improving the accuracy of relevant documents, achieve higher NDCG scores by refining this large set into smaller, more accurately grouped sets. However, it appears that even somewhat less precise sets are more effective for training than a large, unrefined set.

Gemma FT seems to consistently be the best model across all dimensions, confirming the results from the fine-tuning section. Llama FT is not far behind (around 0.001 or 0.002 of NDCG each time), and it would be interesting to leverage the impact of this small difference, as Llama has 1B fewer parameters. In fact, when we have such high NDCG scores, a difference of 0.005 or 0.008 is important—it can truly change the model. However, it is difficult to demonstrate this without directly applying it in a production pipeline.

The MRR metric further supports our analysis, showing a more substantial gain than NDCG. Between the baseline and the best-performing model (Gemma FT), there is a consistent increase of 0.02 to 0.03. This suggests that refining the set of negative documents helps the model better distinguish positive documents and retrieve them in higher positions. As anticipated, the mixed set of negatives—containing both false and true hard negatives with similar relevance scores—can be confusing for the model. For a search engine, achieving the highest possible MRR is essential, as users typically click on the first or second documents. Therefore, these top-ranked documents must accurately fulfill user expectations.

\begin{table}[t]
  \caption{Impact of relevance rescaling methods (None, Llama, Llama FT, Gemma, Gemma FT) on validation dataset ranking metrics (NDCG@k and MRR) across different dimensionality settings, with 95\% confidence intervals from 1000 bootstrap iterations}
  \label{tab:ndcg-performance}
  \centering
  \small 
  \setlength{\tabcolsep}{3pt}
  \begin{tabular}{ll|cccc}
    \toprule
    \textbf{Dim} & \textbf{Rescaling} & \textbf{NDCG@5} & \textbf{NDCG@10} & \textbf{NDCG@30} & \textbf{MRR} \\
    \midrule
    \multirow{5}{*}{768} 
        & None           & 0.952 \tiny{\textcolor[gray]{0.5}{$\pm1\mathrm{e}^{-4}$}} & 0.965 \tiny{\textcolor[gray]{0.5}{$\pm9\mathrm{e}^{-5}$}} & 0.982 \tiny{\textcolor[gray]{0.5}{$\pm5\mathrm{e}^{-5}$}} & 0.865 \tiny{\textcolor[gray]{0.5}{$\pm8\mathrm{e}^{-4}$}} \\
        & Llama          & 0.955 \tiny{\textcolor[gray]{0.5}{$\pm1\mathrm{e}^{-4}$}} & 0.968 \tiny{\textcolor[gray]{0.5}{$\pm9\mathrm{e}^{-5}$}} & 0.984 \tiny{\textcolor[gray]{0.5}{$\pm5\mathrm{e}^{-5}$}} & 0.873 \tiny{\textcolor[gray]{0.5}{$\pm7\mathrm{e}^{-4}$}} \\
        & Llama FT       & 0.958 \tiny{\textcolor[gray]{0.5}{$\pm1\mathrm{e}^{-4}$}} & \textbf{0.970 \tiny{\textcolor[gray]{0.5}{$\pm9\mathrm{e}^{-5}$}}} & \textbf{0.985 \tiny{\textcolor[gray]{0.5}{$\pm4\mathrm{e}^{-5}$}}} & 0.888 \tiny{\textcolor[gray]{0.5}{$\pm7\mathrm{e}^{-4}$}} \\
        & Gemma          & 0.957 \tiny{\textcolor[gray]{0.5}{$\pm1\mathrm{e}^{-4}$}} & 0.969 \tiny{\textcolor[gray]{0.5}{$\pm9\mathrm{e}^{-5}$}} & 0.984 \tiny{\textcolor[gray]{0.5}{$\pm4\mathrm{e}^{-5}$}} & 0.886 \tiny{\textcolor[gray]{0.5}{$\pm7\mathrm{e}^{-4}$}} \\
        & Gemma FT       & \textbf{0.959 \tiny{\textcolor[gray]{0.5}{$\pm1\mathrm{e}^{-4}$}}} & \textbf{0.970 \tiny{\textcolor[gray]{0.5}{$\pm9\mathrm{e}^{-5}$}}} & \textbf{0.985 \tiny{\textcolor[gray]{0.5}{$\pm4\mathrm{e}^{-5}$}}} & \textbf{0.895 \tiny{\textcolor[gray]{0.5}{$\pm7\mathrm{e}^{-4}$}}} \\
    \midrule
    \multirow{5}{*}{64} 
        & None           & 0.942 \tiny{\textcolor[gray]{0.5}{$\pm1\mathrm{e}^{-4}$}} & 0.957 \tiny{\textcolor[gray]{0.5}{$\pm1\mathrm{e}^{-4}$}} & 0.979 \tiny{\textcolor[gray]{0.5}{$\pm5\mathrm{e}^{-5}$}} & 0.831 \tiny{\textcolor[gray]{0.5}{$\pm8\mathrm{e}^{-4}$}} \\
        & Llama          & 0.947 \tiny{\textcolor[gray]{0.5}{$\pm1\mathrm{e}^{-4}$}} & 0.961 \tiny{\textcolor[gray]{0.5}{$\pm1\mathrm{e}^{-4}$}} & 0.980 \tiny{\textcolor[gray]{0.5}{$\pm5\mathrm{e}^{-5}$}} & 0.835 \tiny{\textcolor[gray]{0.5}{$\pm8\mathrm{e}^{-4}$}} \\
        & Llama FT       & \textbf{0.950 \tiny{\textcolor[gray]{0.5}{$\pm1\mathrm{e}^{-4}$}}} & 0.963 \tiny{\textcolor[gray]{0.5}{$\pm1\mathrm{e}^{-4}$}} & 0.981 \tiny{\textcolor[gray]{0.5}{$\pm5\mathrm{e}^{-5}$}} & 0.848 \tiny{\textcolor[gray]{0.5}{$\pm8\mathrm{e}^{-4}$}} \\
        & Gemma          & 0.947 \tiny{\textcolor[gray]{0.5}{$\pm1\mathrm{e}^{-4}$}} & 0.962 \tiny{\textcolor[gray]{0.5}{$\pm1\mathrm{e}^{-4}$}} & 0.981 \tiny{\textcolor[gray]{0.5}{$\pm5\mathrm{e}^{-5}$}} & 0.844 \tiny{\textcolor[gray]{0.5}{$\pm8\mathrm{e}^{-4}$}} \\
        & Gemma FT       & \textbf{0.950 \tiny{\textcolor[gray]{0.5}{$\pm1\mathrm{e}^{-4}$}}} & \textbf{0.964 \tiny{\textcolor[gray]{0.5}{$\pm1\mathrm{e}^{-5}$}}} & \textbf{0.982 \tiny{\textcolor[gray]{0.5}{$\pm5\mathrm{e}^{-5}$}}} & \textbf{0.863 \tiny{\textcolor[gray]{0.5}{$\pm8\mathrm{e}^{-4}$}}} \\
    \midrule
    \multirow{5}{*}{32} 
        & None           &0.930 \tiny{\textcolor[gray]{0.5}{$\pm1\mathrm{e}^{-4}$}} & 0.948 \tiny{\textcolor[gray]{0.5}{$\pm1\mathrm{e}^{-4}$}} & 0.974 \tiny{\textcolor[gray]{0.5}{$\pm6\mathrm{e}^{-5}$}} & 0.773 \tiny{\textcolor[gray]{0.5}{$\pm9\mathrm{e}^{-4}$}} \\
        & Llama          & 0.934 \tiny{\textcolor[gray]{0.5}{$\pm1\mathrm{e}^{-4}$}} & 0.952 \tiny{\textcolor[gray]{0.5}{$\pm1\mathrm{e}^{-4}$}} & 0.976 \tiny{\textcolor[gray]{0.5}{$\pm6\mathrm{e}^{-5}$}} & 0.788 \tiny{\textcolor[gray]{0.5}{$\pm9\mathrm{e}^{-4}$}} \\
        & Llama FT       & 0.937 \tiny{\textcolor[gray]{0.5}{$\pm1\mathrm{e}^{-4}$}} & 0.954 \tiny{\textcolor[gray]{0.5}{$\pm1\mathrm{e}^{-4}$}} & 0.977 \tiny{\textcolor[gray]{0.5}{$\pm5\mathrm{e}^{-5}$}} & 0.793 \tiny{\textcolor[gray]{0.5}{$\pm9\mathrm{e}^{-4}$}} \\
        & Gemma          & 0.935 \tiny{\textcolor[gray]{0.5}{$\pm1\mathrm{e}^{-4}$}} & 0.952 \tiny{\textcolor[gray]{0.5}{$\pm1\mathrm{e}^{-4}$}} & 0.977 \tiny{\textcolor[gray]{0.5}{$\pm5\mathrm{e}^{-5}$}} & 0.789 \tiny{\textcolor[gray]{0.5}{$\pm9\mathrm{e}^{-4}$}} \\
        & Gemma FT       & \textbf{0.938 \tiny{\textcolor[gray]{0.5}{$\pm1\mathrm{e}^{-4}$}}} & \textbf{0.955 \tiny{\textcolor[gray]{0.5}{$\pm1\mathrm{e}^{-4}$}}} & \textbf{0.978 \tiny{\textcolor[gray]{0.5}{$\pm5\mathrm{e}^{-5}$}}} & \textbf{0.809 \tiny{\textcolor[gray]{0.5}{$\pm9\mathrm{e}^{-4}$}}} \\
    \midrule
    \multirow{5}{*}{768 B} 
        & None           & 0.941 \tiny{\textcolor[gray]{0.5}{$\pm2\mathrm{e}^{-4}$}} & 0.953 \tiny{\textcolor[gray]{0.5}{$\pm1\mathrm{e}^{-4}$}} & 0.974 \tiny{\textcolor[gray]{0.5}{$\pm6\mathrm{e}^{-5}$}} & 0.814 \tiny{\textcolor[gray]{0.5}{$\pm9\mathrm{e}^{-4}$}} \\
        & Llama          & 0.943 \tiny{\textcolor[gray]{0.5}{$\pm2\mathrm{e}^{-4}$}} & 0.957 \tiny{\textcolor[gray]{0.5}{$\pm1\mathrm{e}^{-4}$}} & 0.977 \tiny{\textcolor[gray]{0.5}{$\pm6\mathrm{e}^{-5}$}} & 0.825 \tiny{\textcolor[gray]{0.5}{$\pm9\mathrm{e}^{-4}$}} \\
        & Llama FT       & 0.946 \tiny{\textcolor[gray]{0.5}{$\pm2\mathrm{e}^{-4}$}} & 0.959 \tiny{\textcolor[gray]{0.5}{$\pm1\mathrm{e}^{-4}$}} & 0.978 \tiny{\textcolor[gray]{0.5}{$\pm6\mathrm{e}^{-5}$}} & 0.836 \tiny{\textcolor[gray]{0.5}{$\pm9\mathrm{e}^{-4}$}} \\
        & Gemma          & 0.946 \tiny{\textcolor[gray]{0.5}{$\pm2\mathrm{e}^{-4}$}} & 0.958 \tiny{\textcolor[gray]{0.5}{$\pm1\mathrm{e}^{-4}$}} & 0.978 \tiny{\textcolor[gray]{0.5}{$\pm5\mathrm{e}^{-5}$}} & 0.838 \tiny{\textcolor[gray]{0.5}{$\pm9\mathrm{e}^{-4}$}} \\
        & Gemma FT       & \textbf{0.948 \tiny{\textcolor[gray]{0.5}{$\pm2\mathrm{e}^{-4}$}}} & \textbf{0.961 \tiny{\textcolor[gray]{0.5}{$\pm1\mathrm{e}^{-4}$}}} & \textbf{0.979 \tiny{\textcolor[gray]{0.5}{$\pm5\mathrm{e}^{-5}$}}} & \textbf{0.845 \tiny{\textcolor[gray]{0.5}{$\pm9\mathrm{e}^{-4}$}}} \\
    \bottomrule
  \end{tabular}
\end{table}

\section{Discussion \& Future work}
Our approach has some limitations. Firstly, the vast majority of our dataset consists of documents and queries from a French web search engine. Consequently, our methodology may not generalize well to other languages, aside from English. Additionally, we restricted our experiments on LLMs that included French data in their pre-training stage. This limitation highlights the critical role of language-specific training data in achieving high-quality results for relevance assessment in multilingual contexts.


Another limitation lies in our fine-tuning dataset. While small, clean datasets can be powerful, larger and higher-quality datasets enable models to learn and understand tasks more effectively by exposing them to a wider variety of cases.

Additionally, we restricted ourselves to computationally efficient prompting strategies. We leave it to future work to investigate fine-tuning involving advanced techniques such as chain-of-thought (CoT) reasoning, few-shot learning, and others, which could increase model performance at the cost of additional computational resources.


Biases in the LLMs used for labeling may exist that have not yet been identified or addressed. For instance, the models may disproportionately favor documents or websites that were included in their pre-training data, potentially disadvantaging newer or underrepresented sources. Furthermore, content creators may attempt to modify their website content or titles to achieve higher ratings if LLM-based grading is implemented. This would be an important consideration to investigate to guarantee robustness, and the long-term effectiveness of production-ready LLM labeling.



Future research could explore the application of larger LLMs, such as a 70B model (e.g. Llama), with quantization techniques like FP8 or INT8. Quantization offers a compelling trade-off by reducing model size and computational requirements, thereby maintaining or even improving throughput. Given the original model’s higher capacity, this approach holds promise for enhancing performance in relevance labeling and optimizing resource-intensive AI workflows for real-world applications.


\section{Conclusion}
The quality of relevance datasets is a cornerstone of effective ranking model training. Traditionally, the creation of gold-standard labels by human annotators has been costly, time-consuming, and difficult to scale. This paper demonstrates that small fine-tuned LLMs (7B–9B parameters) can generate relevance annotations that closely match the quality of human labels and competitive closed-source models like GPT-4. Our experiments on dense re-ranking tasks reveal that ranker models trained on datasets augmented with small LLM annotations consistently outperform their counterparts trained without this augmentation, paving the way for more scalable, cost-effective, and automated solutions for relevance datasets generation.

\bibliographystyle{ACM-Reference-Format}
\bibliography{refs}

\end{document}